\def\year{2022}\relax
\documentclass[letterpaper]{article} 
\usepackage{aaai22}  
\usepackage{times}  
\usepackage{helvet}  
\usepackage{courier}  
\usepackage[hyphens]{url}  
\usepackage{graphicx} 
\usepackage{natbib}  
\usepackage{caption} 
\DeclareCaptionStyle{ruled}{labelfont=normalfont,labelsep=colon,strut=off} 
\usepackage{algorithm}
\usepackage{algorithmic}

\usepackage{newfloat}
\usepackage{listings}
\floatstyle{ruled}
\newfloat{listing}{tb}{lst}{}
\floatname{listing}{Listing}
\usepackage{enumerate}

\usepackage{amsmath,amssymb,amsfonts}
\usepackage{booktabs}
\usepackage{changes}
\usepackage{cite}
\usepackage{textcomp}

\usepackage{bm}
\usepackage{soul}
\usepackage[normalem]{ulem}

\ifcsname G\endcsname%
    
\else%
    
\fi%
\ifcsname C\endcsname%
    
\else%
    
\fi%

\newcommand{\R}{\mathbb{R}}

\usepackage[all]{xy}

\usepackage{cleveref}
\crefname{section}{§}{§§}
\Crefname{section}{§}{§§}

\newcommand\blfootnote[1]{%
  \begingroup
  \renewcommand\thefootnote{}\footnote{#1}%
  \addtocounter{footnote}{-1}%
  \endgroup
}

\usepackage{lipsum}

\title{Evaluating Automated Driving Planner Robustness against Adversarial Influence}
\author {
   Andres Molina-Markham,
   Silvia G. Ionescu,
   Erin Lanus,\\
   Derek Ng,
   Sam Sommerer,
   Joseph J. Rushanan
}
\affiliations {
    The MITRE Corporation, Bedford, MA, USA\\
    amolinamarkham@mitre.org, jjr@mitre.org
}

\begin{document}

\maketitle

\blfootnote{ {\normalsize Approved for Public Release; Distribution Unlimited. Public Release Case Number 22-0800.
   \copyright 2022 The MITRE Corporation. ALL RIGHTS RESERVED. 
}}

\begin{abstract}
Evaluating the robustness of automated driving planners is a critical and
challenging task. Although methodologies to evaluate vehicles are well
established, they do not yet account for a reality in which vehicles with
autonomous components share the road with adversarial agents. Our approach,
based on probabilistic trust models, aims to help researchers assess the
robustness of protections for machine learning-enabled planners against
adversarial influence. In contrast with established practices that evaluate
safety using the same evaluation dataset for all vehicles, we argue that
adversarial evaluation  fundamentally requires a process that seeks to defeat a
specific protection. Hence, we propose that evaluations be based on estimating
the difficulty for an adversary to determine conditions that effectively induce
unsafe behavior. This type of inference requires precise statements about
threats, protections, and aspects of planning decisions to be guarded. We
demonstrate our approach by evaluating protections for planners relying on
camera-based object detectors.
\end{abstract}

\section{Introduction}\label{intro}

As autonomous vehicle (AV) technology matures towards higher SAE automation
levels and intended human intervention decreases, unexpected behavior by
automated driving planners can cause increasingly devastating effects. AVs
utilize  object detectors developed via machine learning (ML) for planning
tasks, such as the decision to brake due to the perception of a stop sign.
Discovery of attacks against object detectors indicates that unsafe behavior can
occur not only as the result of rare natural events, but due to adversarial
\emph{disturbances}, deliberate actions by an adversary. Trustworthy
\emph{protections} aimed at preventing or reacting to adversarial influence
are needed. Trusting protections to guard the decisions of a planner requires an
evaluation methodology to: \textbf{compare protections} to unambiguously
determine which is better for mitigating a specific threat; and, \textbf{justify
confidence} in a protection to determine how well it guards a planner's
decisions against specific adversaries.

The robotics and automation research community understands the criticality of
evaluating safety for AVs~\cite{kone_safety_2019}. Moreover, recent guidance
recognizes special considerations for the safe use of ML in autonomous
systems~\cite{hawkins2021guidance}. Nevertheless, current practices and
recommendations do not include satisfactory methodologies to evaluate
protections for ML-enabled  automated driving planners, a problem at the
intersection of AV safety, AV security, disturbance generation via simulation,
and adversarial ML.

Guidance on the assurance of ML-enabled autonomous systems recommends evaluation
datasets that are relevant, complete, and balanced~\cite{hawkins2021guidance}.
The call for complete evaluation datasets already poses a challenge in
non-adversarial situations due to the long tail of rare cases in the complex
environments in which AVs operate. The space of possible tests is larger when
including disturbances due to adding the actions of an adversary to the other
parameters. Thus, a static test set cannot be expected to feasibly cover the
space for adversarial evaluation. Additionally, adversarial actions occur
strategically rather than by chance, and consequently, once effective
disturbances are discovered, they cannot be considered rare corner cases.  

Existing work has already argued that benchmarks do not adequately assess risk
across deployment spaces~\cite{norden2019efficient}. The safety research
community relies on simulation to evaluate AVs when complexities of systems and
environments are too difficult to fully model~\cite{kalra2016driving,
corso2021survey} and to generate disturbances via search techniques like
Bayesian Optimization (BO) \cite{tuncali2018simulation,
abeysirigoonawardena2019generating, nguyen2020bayesian}. Work in AV security is
broader in scope and typically includes other more traditional aspects of
security shared with cyberphysical systems~\cite{thing2016autonomous} and
software assurance~\cite{chattopadhyay2020autonomous}. 

Work in ML adversarial robustness  \cite{szegedy2014intriguing,
nicolae2019adversarial} has brought awareness of the susceptibility of ML
algorithms to a wide range of attacks. However, system evaluations need context
of how ML algorithms are used and protected in specific applications. For
example, a defense for object detectors~\cite{braunegg2020apricot} using
adversarial training is not evaluated in the context of how a planner will use
the object detector's decisions. In general, works related to algorithm attacks
such as adversarial physical attacks~\cite{brown2018adversarial} and adversarial
policies~\cite{gleave2019adversarial} are complementary to our work. Currently,
however, AV planners do not implement protections inspired by algorithm
defenses---e.g., \cite{salman2020denoised, braunegg2020apricot,
abeysirigoonawardena2019generating, gleave2019adversarial,
thing2016autonomous}--- against these adversaries.

We argue that the effectiveness of protections can only be adequately evaluated
when the assumptions about threats and valid behavior to preserve are well
specified. This is key to be able to compare protections against a specific
threat. Comparison is necessary, but not sufficient. Even with a suitable
solution, it is possible to draw inaccurate conclusions about the overall
robustness of a protection. For example, a fixed set of tests may suggest a
protection has better performance than another, yet an adversary could more
easily find new tests that are effective against the seemingly stronger
protection. The strategy for selecting the tests for each protection must also
be evaluated. Thus, to justify confidence in a protection, we need to assess the
difficulty for an adversary to find the parameters of an effective disturbance
over an infinite, but well-defined, set of possible disturbances.

To address current limitations, the contribution of this work is to propose a
framework combining established ideas about threat modeling to engineer secure
systems \cite{shevchenko_threat_2018} with state-of-the-art ideas about
probabilistic knowledge representation and inference to evaluate protections for
AVs. This framework employs \emph{probabilistic trust models (PTMs)} to state
probabilistic assumptions about adversarial threats, protections, and aspects of
valid AV planner behavior that must be preserved. Encoding PTMs via
\emph{generative programs}~\cite{cusumano-towner_gen:_2019} enables inference
tasks necessary for estimating effectiveness of a disturbance against a
protection. Together, generative programs and simulation enable computational
estimates of the risk posed by a disturbance from the prior probability
distribution over the set of disturbances and the posterior probability of
faults induced determined by running the disturbance in simulation. We improve
upon simulated disturbance generation via search by defining assurance metrics
using probabilistic trust models that more adequately encode priors and
inference tasks related to concrete definitions of valid behavior to preserve.
Hence, our work offers a way to contextualize the work of multiple communities,
including the adversarial ML robustness community, to concretely evaluate the
effectiveness of a set of protections for a specific planner. Our tool, PARAPET,
automates the generation of synthetic data using PTMs encoded in generative
programs combined with simulation to justify confidence in protections for an
ML-enabled automated driving planner.

\section{Method}\label{method}

The \emph{falsification safety validation task}~\cite{corso2021survey} seeks
failure examples and computes search coverage metrics to reason about confidence
in the safety of autonomous cyber-physical systems. PARAPET extends
falsification to failures caused by adversarial threats. When a system and an
adversary interact with an environment, a safety property $\psi$ is defined over
(ground truth) state sequences $\boldsymbol{s}=[s_1,...,s_t]$, where $s_t \in
S$ is the state of the environment, at time $t \in \{1,...,t_{max}\}$. A
\emph{disturbance} is an action $x$ from the adversary. The falsification task
is defined as the process of finding a disturbance sequence $\boldsymbol{x}$ for
$\boldsymbol{s}$ that induces a state sequence $\boldsymbol{s'}$ which violates
the safety property $\psi$ for the system equipped with the protection. That is,
$\boldsymbol{x}$ is a \emph{counterexample} demonstrating that $\psi$ does not
hold. Assurance provided by the protection is related to the difficulty of
finding a counterexample. To perform the falsification task, it is necessary to
(1) turn the task into an optimization problem over disturbance sequences
with a suitable objective function; and (2) attempt to solve the problem
efficiently with appropriate coverage metrics.

The \emph{objective function} $f: X \mapsto \R$, $\boldsymbol{x} \in X$  such
that, for a safety threshold $\epsilon, f(\boldsymbol{x})< \epsilon$ (dependent
on risk tolerance)  if and only if  the probability that $\boldsymbol{x}$  is a
counterexample is above an assurance threshold $\delta$.
\citeauthor{corso2021survey}'s objective functions are designed such that
$f(\boldsymbol{x})< \epsilon$  if and only if $\boldsymbol{x}$ is a
counterexample. In practice, it may not be possible to state with certainty that
a specific disturbance $\boldsymbol{x}$ is a counterexample when the effect of
$\boldsymbol{x}$ is non-deterministic or depends in part on aspects of the
environment or system. In contrast, PARAPET objective functions are encoded via
PTMs, which define a distribution over possible outcomes (cf. \emph{traces of
generative programs} below) of a simulation. A simulation returns the observed
sequence $\boldsymbol{r}$ of state representations (e.g., sensor readings)
corresponding to $\boldsymbol{s}$ disturbed by $\boldsymbol{x}$. PARAPET uses
$\boldsymbol{r}$ to estimate the probability that $\boldsymbol{x}$ will induce
$\boldsymbol{s'}$ to violate $\psi$. To do so, it is neither sufficient nor
necessary to observe a failure; a failure can be caused by some other factor
than $\boldsymbol{x}$, and $\boldsymbol{x}$ may not cause a failure in every run
due to non-deterministic factors. Therefore, as opposed to
\citeauthor{corso2021survey}'s deterministic mapping between $\boldsymbol{x}$
and $\boldsymbol{s}$, we think of the set of sequences $S$ that are not only
induced by $\boldsymbol{x}$, but also by actions of the system and uncertain
aspects of the environment.

\begin{figure}
    \centering
      \includegraphics[width=.4\textwidth]{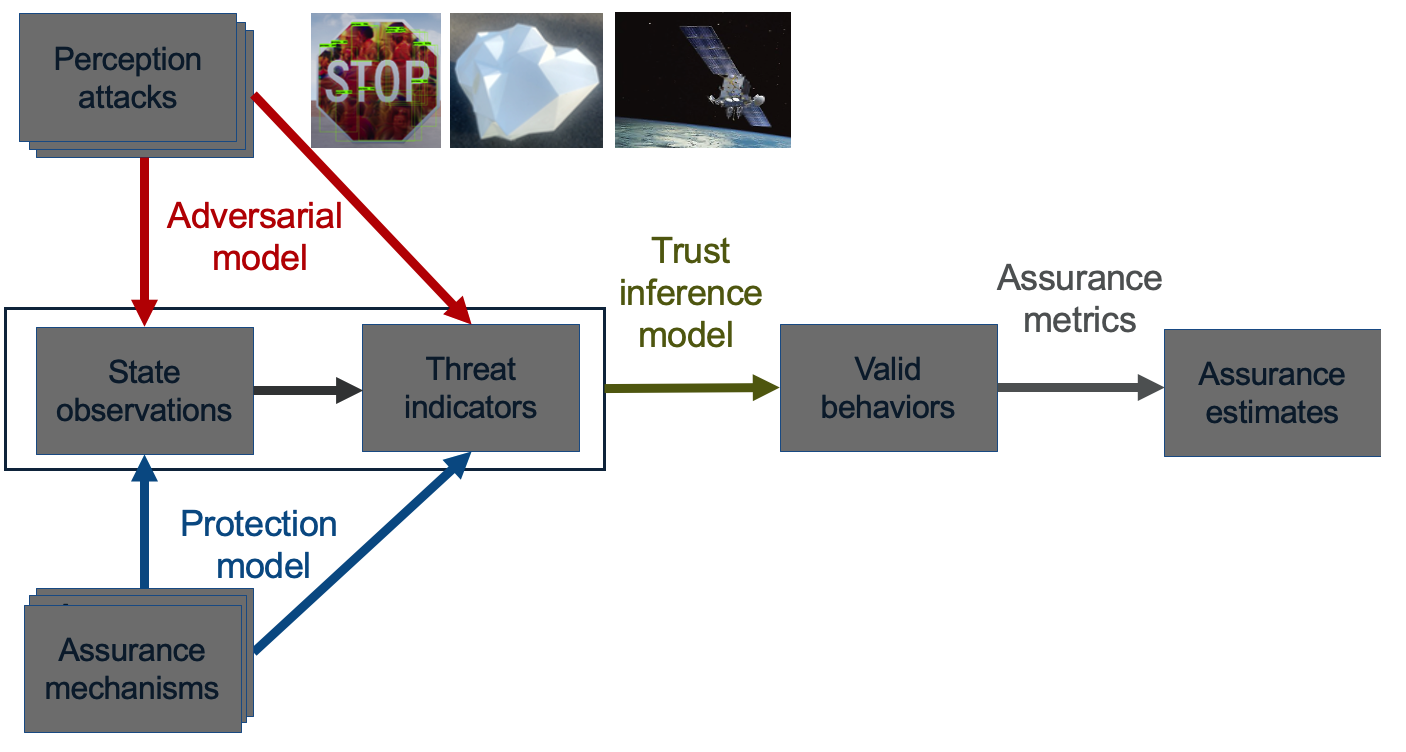}
    \caption[Probabilistic Trust Models]{PTMs model 
    relationships between threats, protections, and 
    behavior to reason about assurance.}
    \label{fig:trust_models}
\end{figure}

In PARAPET testers encode safety properties and assumptions via PTMs specified
via generative programs in languages such as
Gen~\cite{cusumano-towner_gen:_2019}. PTMs define probability distributions on
traces of programs ($\omega,\zeta$) that can make probabilistic random choices
during execution where $\zeta$ is the weight proportional to the likelihood of a
program trace $\omega$. When a generative program $P$ is executed with a
parameter vector $\boldsymbol{z}$, it produces the trace $\omega$ with
probability $p(\omega;P,\boldsymbol{z})$, which is useful to drive simulations
subject to the assumptions in the adversarial model. Generative programs also
enable the automatic estimation of posterior distributions $p(\omega
|\boldsymbol{y};P,\boldsymbol{z})$, associated with the objective function for
the falsification task. When these models are conditioned on observations, they
induce posterior distributions that allow computation of assurance metrics. 

Hence, PTMs relate threats, protections, and aspects of valid behavior to be preserved,
and are typically composed of an adversarial model, a protection model, a trust
inference model, and a set of assurance metrics (Figure~\ref{fig:trust_models}).
\emph{Adversarial models} express prior assumptions about the likelihood of
disturbances, the probable effects of disturbances on observed state
representations, and the constraints of the adversary, such as whether the
adversary has white or black box knowledge of the system, including the
protection. \emph{Protection models} specify how protections affect observed
state representations via assumptions about the state prior to applying and
limitations on the effect of the protection. Protections can passively attempt
to detect the presence of an adversary or can actively modify observations to
mitigate possible disturbances. \emph{Trust inference models} specify how to
infer valid behavior from observations that could have been influenced by
disturbances and protections. The posterior distributions also reflect
uncertainties associated with making decisions with partial and imperfect
information. Trust inference models take into account how actions selected by a
planner may result in unsafe conditions if, for example, an object were
misdetected, or an estimated location were inaccurate. We use trust inference
models to implement objective functions (cf. \cref{use case}). \emph{Assurance
metrics} capture (i) the objective function measuring the effectiveness of a
disturbance over a state sequence in causing invalid behavior (by summarizing
the posterior probability distributions via numeric degrees of belief associated
with the probability that a planner's decisions will result in invalid
behavior); and (ii) an estimate of adversary difficulty in finding an effective
disturbance (by describing coverage notions related to the search space of
disturbances). Types of coverage metrics include probabilistic,
disturbance-space, and state-space coverage, which we leave to future work.

PARAPET combines PTMs with a strategy to explore the set of disturbances guided
by the objective function to solve the associated falsification task
efficiently. While an adversary can employ any strategy, solving the
optimization problem via an uninformed search strategy is not feasible in
general. The best known informed strategies either rely on gradient methods when
$f(\boldsymbol{x})$ is differentiable and inexpensive to evaluate or BO when
$f(\boldsymbol{x})$ is a black-box function that can only be noisily and
expensively evaluated. PARAPET returns any counterexamples discovered in
simulation, disturbances $\{\boldsymbol{x}\}$ found to cause the system to
violate $\psi$. If none are found, PARAPET returns a dataset
$\{\boldsymbol{x_i}, \boldsymbol{r_i}\}$ of pairs and an estimate of the number
of attempts the adversary needs to perform to find a counterexample, if the
adversary used a similar strategy to solve the falsification task.

\section{Use Case}\label{use case}

PARAPET is illustrated by evaluating a \emph{sensor-fusion} protection designed
for automated driving planners relying on camera-based object detectors. Valid
behavior requires the planner to initiate a transition to a minimal risk
condition, alerting the driver and disengaging autonomous driving, when the
protection predicts the presence of a disturbance.

The adversary is assumed to not have internal access to the system and is unable
to directly manipulate digital data, but can create static physical artifacts
via the ShapeShifter attack~\cite{chen_shapeshifter:_2018}. 

ShapeShifter computes a perturbation that is physically applied to an object to
induce misclassifications effective in multiple bounding boxes detected at
different scales, viewing angles, and distances over a range of lighting
conditions and camera constraints. It is demonstrated in the original paper on
the Faster-RCNN object detector~\cite{ren_faster_2015}. Because the attacked
object detector is easily available, the adversary is assumed to have
``white-box" knowledge about the ML model structure and weights. To model an
adversary with practical computational powers, the perturbation need not be
computed in real time. The adversary is assumed to only be able to manipulate
certain regions of an object and cannot change the shape of an object, such as
the octagonal shape of a stop sign. The perturbation \emph{strength} corresponds
to ShapeShifter's hyperparameter $c$. The adversary can manipulate the placement
of the object in multiple dimensions relative to its standardized location. A
disturbance is defined via five continuous parameters: the perturbation strength
applied to and the placement of a stop sign as deviation from the normal height,
roll, pitch, and yaw. As the AV operates in a world with humans, disturbances
are practically limited in strength and deviation from the norm. Large
disturbances causing humans to misclassify objects may be replaced by a
different object entirely, nullifying use of the ShapeShifter attack.
Conspicuous disturbances that draw the attention of humans are detected, unable
to cause unexpected effects (i.e., this is related to minimizing the
\emph{strength} of a perturbation while maintaining attack effectiveness (cf.
~\cite{chen_shapeshifter:_2018})). The adversary's goal is to induce invalid
actions by reducing the planner's confidence that it is detecting a stop sign in
time to brake.

The sensor-fusion protection model (implemented via a generative program
$P_s$) encodes probability distributions of traces $p(\omega_s;P_s,\boldsymbol{z_s})$, where
$\boldsymbol{z_s}$ corresponds to sensor data. That is, $P_s$ encodes a prior distribution of
correlations between object detectors, derived from simulations without an
adversary. $P_s$ also encodes an assumption that when an adversary is present,
the correlations between object detectors are drawn from a uniform distribution.
The protection also encodes an assumption about the existence of disturbances
relative to the location of real stop signs using Poisson distributions. This
way, $P_s$ integrates streams of observations from three sensors, Faster-RCNN
and Yolo3~\cite{redmon_yolov3_2018} object detectors and GPS, as well as mapping
information giving the expected locations of stop signs.  
 
$P_s$ can be used to estimate the posterior $p(\omega|\boldsymbol{y_s};P_s,\boldsymbol{z_s})$ conditioned
on new observations $\boldsymbol{y_s}$. When likely traces $w$ (consistent with this
posterior) correspond to traces where an adversary may be present, the
sensor-fusion protection raises an alert. This protection is compared to a
\emph{trivial} protection that never alerts.

The objective function evaluates the effectiveness of a disturbance to evade alerts in
 a situation that may likely lead to an accident.  
The disturbance is effective if the protection does not alert when
$\boldsymbol{x}$ is present and $\boldsymbol{s'}$ violates $\psi$ (such that the
vehicle does not detect the attacked stop sign). The disturbance is moderately
effective when the alert is raised outside of possible braking distance or
within possible but unsafe braking distance given the speed. The disturbance is
mildly effective when it results in unnecessary alerts, i.e.,  asking the driver
to take control unnecessarily when the planner is likely to adequately detect
the stop sign and stop as a result. Last, the objective function favors
disturbances that are less likely to be noticed (relative to the deviation from
the norm w.r.t. placement and strength of the perturbation).

\section{Experiments}\label{experiments}

\begin{figure}[t]
  \centering
    \includegraphics[width=.35\textwidth]{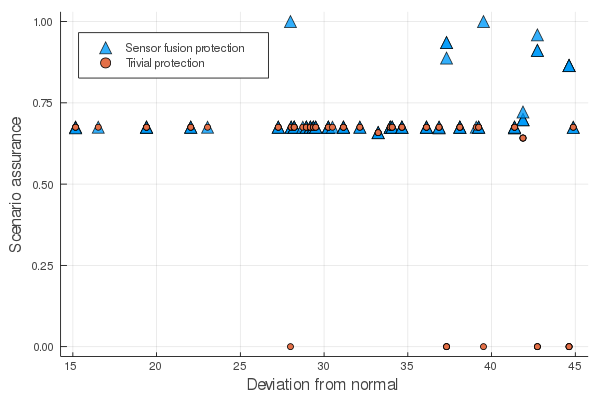}
  \caption[Comparison]{Comparison between trivial and sensor-fusion
  protections plotting Euclidean distance
  to an undisturbed sign 
  against objective function score.}
  \label{fig:comparison}
\end{figure}

\begin{figure}[t]
  \centering
    \includegraphics[width=.35\textwidth, trim={0 21 0 19},clip]{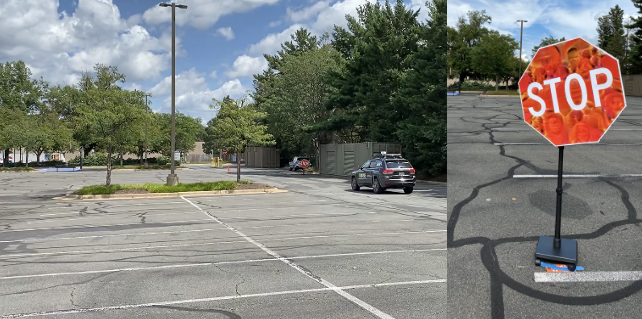}
  \caption[Validation]{Validation setup.}
  \label{fig:validation}
\end{figure}

\begin{table}[t]
\centering
\begin{tabular}{l r r r r | r | r}
\toprule
Strength	&	Height	&	Roll	&	Pitch		& 	Yaw		&	Sim  & Real \\
\midrule
0.0077		&	0.36'	&	$5^{\circ}$			&	$22^{\circ}$			&	$-23^{\circ}$			&	0.68 & 0.82\\
0.0100		&	-0.48'	&	$-12^{\circ}$		&	$-3^{\circ}$			&	$-23^{\circ}$			&	0.70 & 0.87\\
0.0050		&	-0.05'	&	$22^{\circ}$		&	$-22^{\circ}$			&	$5^{\circ}$			&	1.00 & 1.00\\
\bottomrule
\end{tabular}
\caption{Parameters of min, mean, and max scores. \emph{Strength} is ShapeShifter's $c$ confidence
parameter~\cite{chen_shapeshifter:_2018}. Smaller values of this parameter
result in more conspicuous but also more robust attacks. \emph{Height},
\emph{Roll}, \emph{Pitch}, and \emph{Yaw}, correspond to the placement of a stop
sign. \emph{Sim} corresponds to scores computed from simulations, and
\emph{Real} corresponds to scores computed using a real vehicle (cf. ~\cref{fig:validation}).}
\label{test2scenarios}
\end{table}

We demonstrate PARAPET's ability to compare two protections against a specific
adversary and measure the effectiveness of a disturbance---towards reasoning
about the difficulty for an adversary to defeat a protection. We demonstrate via
real-world validation experiments that PARAPET's simulation results reasonably
approximate evaluation of a protection on a real vehicle against a real
adversary.

All simulations are performed using CARLA's Scenario
Runner~\cite{dosovitskiy2017carla} with a single camera placed on the front of
the vehicle. The priors to inform the protection were derived from comparing the
detections of Faster-RCNN and Yolo3 on observations from  
45 state trajectories exercising different weather and road conditions. For the
simulation experiments, we used a bootstrapping set with 100 trajectories to
initialize BO search. Figure~\ref{fig:comparison} plots protection scores
against trajectories, demonstrating that the objective function distinguishes
between the sensor-fusion and trivial protection on the disturbances where
sensor-fusion alerts while the trivial one does not. When the disturbances are
large, the sensor-fusion protection detects the disturbance more often as
indicated by more scores above the average, but the score is also penalized for
becoming increasingly obvious to humans as indicated by the slight downward
trend. Last, both protections receive similar scores for a subset of
disturbances. If this subset is used as a fixed test set, performance of the two
protections is indistinguishable, emphasizing that the strategy for selecting
tests is crucial. For that, we use an approach similar to
\cite{abeysirigoonawardena2019generating} to search the disturbance space
intending to solve the falsification task. After 150 iterations, BO with an
expected improvement acquisition function fails to find a disturbance that
induces invalid behavior. Scores are similar to Figure~\ref{fig:comparison}.
Each simulation took about one hour of computation, totaling over 10 days for
both experiments, providing insight into the difficulty of defeating the
sensor-fusion protection. 

Real world validation experiments were conducted using a vehicle outfitted with
a Blackfly S GigE Machine Vision Camera, a KVH GEO FOG 3D Dual GNSS/INS in an
urban parking lot at midday with partly cloudy weather. Simulations were
parameterized to match the state sequence for the test environment to search for
effective disturbances. Perturbations were printed to scale as illustrated in
Figure~\ref{fig:validation}.   
Potential for discrepancies between simulation and real-world is due to
approximation error in positioning or print medium inaccuracy. The tests
selected for validation in Table~\ref{test2scenarios} represent disturbances
with low, average, and high scores. The validation scores  are not identical but
are ranked the same, suggesting that disturbances that are effective in
simulation are likely effective in practice.

\section{Conclusions}\label{conclusions}

PARAPET is an approach to evaluate protections for automated driving planners
against adversarial influence. Generative programs model assumptions about the
valid behavior that the protections must preserve under a concrete threat model.
We validate that simulation results have real-world applicability. We
demonstrate that evaluations using fixed test sets are not sufficient for
justifying confidence in a protection; it is also necessary to reason about the
sampling strategy. In future work, we are exploring coverage metrics to reason
about the number of evaluations needed to quantify the difficulty of finding
disturbances. The protection we evaluate is passive, but algorithmic defenses to
ML attacks can be actively applied to mitigate threats. Denoised smoothing
guarantees that an image classifier is $l_p$-robust to adversarial examples
\cite{salman2020denoised}. We are evaluating the extent to which this may be
adapted to make the Faster-RCNN object detector in the sensor-fusion protection
robust. Preliminary results suggest denoised smoothing lowers accuracy at a
different rate for different parameters, underscoring the usefulness of PTMs for
evaluating protections. We are also using assurance metrics to implement
planner protections as online monitors.

\section*{Acknowledgements}
We thank Zach LaCelle and members of the \emph{Mobile Autonomous Systems
Experimentation Lab} for their assistance with experiments with an automated
vehicle. We also thank Derek G. Tsui for his work improving PARAPET.

\bibliography{auto_references, DAPSPAC}

\end{document}